\documentclass{article}
\usepackage[T1]{fontenc} % add special characters (e.g., umlaute)
\usepackage[utf8]{inputenc} % set utf-8 as default input encoding
\usepackage{ismir,amsmath,cite,url}
\usepackage{graphicx}
\usepackage{color}

\usepackage{placeins}
\usepackage{lipsum}
\usepackage{subcaption}
\usepackage{cleveref}
\usepackage{stfloats}
\usepackage{soul}
\usepackage{multirow}

\newcommand{\shsfive}{SHS$_\text{5+}$}
\newcommand{\shsfour}{SHS$_\text{4-}$}
\newcommand{\shsfivefour}{SHS$_\text{5+/4-}$}

\usepackage{lineno}
\title{And what if two musical versions don't share \\ melody, harmony, rhythm, or lyrics ?}
\twoauthors
  {Mathilde Abrassart} {Ircam Amplify\\ {\tt abrassart@ircam.fr}}
  {Guillaume Doras} {Ircam, Sorbonne Université, CNRS, STMS Lab\\ {\tt doras@ircam.fr}}

\def\authorname{M. Abrassart, G. Doras}

\begin{document}

\maketitle
\begin{abstract}
%The abstract should be placed at the top left column and should contain about 150-200 words.

Version identification (VI) has seen substantial progress over the past few years. On the one hand, the introduction of the metric learning paradigm has favored the emergence of scalable yet accurate VI systems. On the other hand, using features focusing on specific aspects of musical pieces, such as melody, harmony, or lyrics, yielded interpretable and promising performances. In this work, we build upon these recent advances and propose a metric learning-based system systematically leveraging four dimensions commonly admitted to convey musical similarity between versions: melodic line, harmonic structure, rhythmic patterns, and lyrics. We describe our deliberately simple model architecture, and we show in particular that an approximated representation of the lyrics is an efficient proxy to discriminate between versions and non-versions. We then describe how these features complement each other and yield new state-of-the-art performances on two publicly available datasets. We finally suggest that a VI system using a combination of melodic, harmonic, rhythmic and lyrics features could theoretically reach the optimal performances obtainable on these datasets.

\end{abstract}

\section{Introduction}\label{sec:introduction}

The version identification (VI) problem has received much attention over the last two decades. Pioneering works showed promising accuracy on small audio datasets, but remained difficult to scale to larger modern audio corpora. The recent introduction of data-driven approaches based on neural networks led to significant progress towards accurate yet scalable VI systems \cite{yesiler2021audio}.

Different paradigms are currently active: one approach considers VI as a classification task, and intends to classify versions into the same class \cite{yu2019learning}, while another approach formulates VI as a metric learning problem, and intends to minimize (resp. maximize) a distance between versions (resp. non-versions) \cite{doras2019cover, yesiler2019accurate}. Recent works have also proposed a combination of both \cite{du2022bytecover2}. Metric learning or classification approaches seem to yield similar performances, as it has been observed for other MIR applications \cite{lee2020disentangled}. These systems also differ according to a perhaps more important aspect: their input feature. Some use a generic audio representation, such as the Constant-Q transform \cite{yu2019temporal}, and rely on the expressivity of the network to disentangle relevant musical features. Others use specialized features, such as the melodic line, the harmonic structure and/or the lyrics, and rely on input data to discriminate between versions and non-versions \cite{doras2019cover, yesiler2019accurate, doras2020combining, vaglio2021words}.

In this work, we pursue in the direction of a metric-based approach using specialized features. % NEW
We build upon the system described in \cite{doras2020combining}, conserving its principle and architecture, and explore the use of new input features. %
The reason is motivated by three practical considerations: firstly, the metric learning approach yields a very compact representation of audio (its embedding), which can be conveniently stored, indexed and queried in very large databases. Secondly, the embedding space and the musical similarity measure that is obtained for each different specialized features is meaningful from a musical perspective, and can be reused for other purposes, e.g. playlist generation. Thirdly, the use of specialized features requires smaller models that are faster to train and %thus are more frugal from an energy consumption perspective \cite{douwes2021energy}.
less energy consuming than larger architectures \cite{douwes2021energy}.

It can reasonably be assumed that different versions of the same musical work share at least one of these four features: the melodic line, the harmonic structure, the rhythmic patterns and sometimes the lyrics. The role of melody and harmony in version similarity has been thoroughly investigated \cite{doras2020combining}. The role of the lyrics has also been studied recently, albeit not from a metric learning perspective \cite{vaglio2021words}. 

In this work, we present a systematic study of the contribution of these four features to version similarity, and describe a metric learning-based system combining all of them. We show that this combination provides new state-of-the-art performances on two publicly available datasets. %We also show that the information contained in the combined feature embeddings is theoretically sufficient to build a VI system with perfect accuracy.
We also show that an oracle using these feature embeddings nearly achieves the maximum theoretical performances on these datasets, suggesting that design of future VI systems reaching these performances may be possible. %NEW GD

The rest of this paper is organized as follows: we briefly review the previous studies inspiring our current work (\secref{sec:related_work}). We then describe how we extracted rhythmic and lyrics features, and our metric learning-based architecture (\secref{sec:our_method}). We present our experiments, discuss our results (\secref{sec:quantitative}), and illustrate them with some examples (\secref{sec:qualitative}). We conclude this paper with our future work.

\section{Related work}
\label{sec:related_work}

In this section, we present a brief overview of the main concepts that inspired our present work.

\subsection{Metric learning}

Learning a similarity metric that generalizes to unseen examples is a common objective in machine learning. The goal is to learn how to map the data of interest into a compact representation (its embedding), and to minimize (resp. maximize) the distance between the embeddings of similar (resp. dissimilar) examples. Various MIR applications rely on a concept of musical similarity, e.g. music classification \cite{slaney2008learning}, music recommendation \cite{mcfee2012learning}, or VI systems \cite{yesiler2021audio}, among many others. Musical similarity between two tracks is typically evaluated first deriving an intermediate feature representation from the audio waveform and then computing a distance between feature pairs. 

In the past few years, metric learning has proven its efficiency to build scalable yet accurate VI systems. These modern architectures typically rely on a CNN-based model trained with a triplet loss  \cite{schroff2015facenet} to embed the musical information contained in the input feature into a single vector embedding that can be rapidly compared via Euclidean distance computation. However, these different systems have made different choices regarding their input features: for instance, Doras et al. \cite{doras2019cover} used a melodic line representation, Yesiler et al. \cite{yesiler2019accurate} used a harmonic structure representation, and Du et al. \cite{du2022bytecover2} used a more generic CQT. The choice between specialized or generic features seems to have a non-negligible impact on the required size of the models (the former uses a 5-layers CNN, while the latter uses a ResNet50).

In this work, we choose the first alternative, and we propose to explore other specialized features beside melody and harmony, in particular the rhythmic and lyrics features.

\subsection{Rhythm patterns detection}
\label{subsec:fp}
Musical similarity based on rhythmic patterns has long been investigated in MIR research, e.g. for audio retrieval \cite{foote2002audio} or music classification \cite{dixon2004towards}. With the purpose of analysis of musical style and recognition of musical genres, Pampalk et al. introduced the fluctuation patterns (FP), representing rhythmic patterns in different frequency bands, and their evolution over time \cite{pampalk2006computational}. The basic assumption is that similar songs exhibit similar characteristic rhythmic patterns, and that comparing FP between tracks shall give enough information about their similarity or dissimilarity from a genre or mood perspective. We propose here to extend this idea to VI context, and to use the fluctuation patterns to discriminate versions from non-versions.

In practice, the FP is obtained computing the audio Mel spectrogram, summing up the high Mel bands to highlight the low frequencies and performing a second STFT in each mel band along the time axis. This results in a 3-dimensional matrix with axes corresponding to the the Mel bands, the frequency modulation and the time. The frequency modulation axis therefore represents the periodicity of the loudness in the corresponding Mel band: for example, a drum kick playing at 120 bpm will be represented here with a frequency modulation at 2 Hz in the low frequency bands. Finally, a perceptual filter is applied on each frequency modulation band. This filter is supposed to highlight the frequency modulations most perceived by the human ear ; for example, a frequency modulation at 4 Hz gives a more intense feeling of fluctuation strength. The 3-dimensional matrix is then averaged along the frequency axis, resulting in a representation of the variations of the frequency modulation over time.  

One of the FP limitation is the use of a linear scale to represent periodicities. This was addressed by Pohle et al., who used a log scale to represent frequency modulations \cite{pohle2009rhythm}. The advantage is that the same onset structure played at different tempi will have all its activations shifted by the same amount along the frequency modulation axis. Another way to achieve this tempo invariance is to compute periodicities with a Constant-Q transform (CQT) instead of a STFT \cite{foroughmand2019deep}.

\subsection{Lyrics recognition}
In automatic speech recognition (ASR), the traditional approach relies on a language model and an acoustic model, typically implemented as a Hidden Markov Model, possibly coupled with a neural network \cite{trentin2001survey}. An alternative approach consists in implementing both language and acoustic models as a single neural network, trained in an end-to-end fashion with a Connectionist Temporal Classification (CTC) loss \cite{graves2006connectionist}. CTC-based models outputs the probability distribution of symbols at each time frame, which can be decoded into the most likely sequence of symbols via classical beam search. This approach has become very popular since fully convolutional end-to-end architectures have achieved performances comparable to those of the hybrid architectures \cite{collobert2016wav2letter}. 

Although singing voice has many obvious differences with speech, automatic lyrics recognition (ALR) or alignment (ALA) systems are usually directly inspired by ASR applications. Moreover, the recent introduction of large lyrics annotated audio datasets, such as Dali \cite{meseguer2018dali}, has fostered the development of new ALR systems, whether they are based on the traditional \cite{gupta2019acoustic} or end-to-end \cite{stoller2019end} architectures. While the former seems to yield better results \cite{gupta2020automatic}, the latter has the advantage of its simplicity, both at training and inference time.

However, very few attempts have been made to use lyrics to assess audio track similarity. It was proposed to improve a query-by-humming system \cite{wang2014improving}, but to the best of our knowlege, only Vaglio et al. proposed the use of lyrics for version identification \cite{vaglio2021words}. They used an existing ALR system to extract lyrics from the audio, and estimate track similarity via a string matching algorithm. However, the comparison cost of such algorithm quickly becomes prohibitive when querying large modern corpora, and limits the scalability of this approach.

\begin{figure*}[t]
    \begin{center}
    \includegraphics[width=1\textwidth]{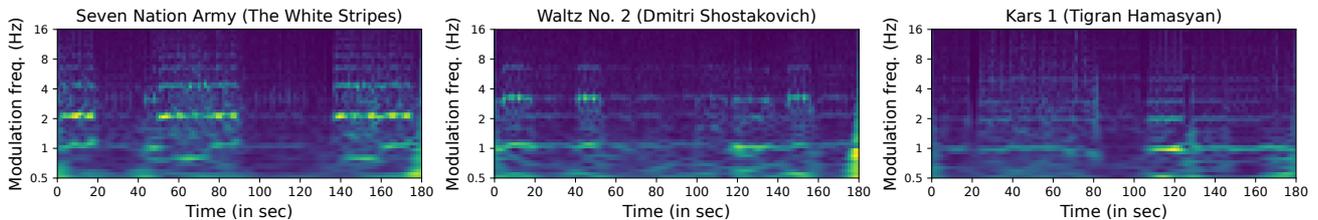}
    \vspace{-1.5em}
    \caption{Examples of the Constant Q Fluctuation Patterns (CQFP).}   
    \label{fig:fp_example}
    \end{center}
%\vspace{-1.5em}
\end{figure*}

\section{Proposed method}
\label{sec:our_method}

In this section, we describe and motivate our design choices. We first present how we extracted our rhythmic and lyrics features, and publicly release our datasets\footnote{\label{note1}https://ircam-anasynth.github.io/papers/2022/abrassart}. We then present our metric learning-based VI model. % NEW
%\vspace{-0.3 cm}
\subsection{Rhythmic features}
\label{subsec:rhythmic_features}

Our assumption is that the FP representation described in \secref{subsec:fp} displays both the local rhythmic patterns along the frequency modulation axis, as well the global rhythmic evolution of the piece over time. We therefore propose to use this representation as our rhythmic feature for version identification. However, we introduce a CQT to compute the periodicities, and to achieve tempo invariance along the frequency modulation dimension. We choose as minimum frequency for the CQT 0.5 Hz (i.e. 30 bpm which we assume would be the tempo of the slowest tracks), and to cover up to 5 octaves i.e. 16Hz (960 bpm) with 10 bins per octave to get all the rhythmic subtleties. We kept the same other parameters as in the original implementation \cite{pampalk2006computational}. 

As an illustration, \figref{fig:fp_example} represents the CQ-FP obtained for different tracks with characteristic time signature. The first example has a 4/4 time signature, and clearly displays a rhythmic pattern present around 2 Hz (\textasciitilde120 bpm), as well as another periodicity around 4 Hz, which corresponds to a clear binary rhythm. The second example has a 3/4 time signature, and the rhythm of a waltz appears clearly: one beat at 1 Hz (60 bpm) and another at about 3 Hz. Finally, the third example has a 5/4 time signature (irregular), and we find this characteristic by observing bands at 1, 2 and 3 Hz. It can also be observed on each example that our rhythmic feature also represents the global structure of the piece over time, which is probably another relevant aspect in a VI context. Finally, as the modulation frequency dimension has a constant Q-factor, a change in tempo would not change the spacing between activations.

\subsection{Lyrics features}
\label{subsec:lyrics_features}

We argue that accurate lyrics recognition is not required for version identification, and that identifying only a few common words, or even a few common character sequences, between tracks shall be sufficient to determine whether they are versions or not. We thus implemented a deliberately simple fully convolutional ALR system inspired by recent ASR system \cite{collobert2016wav2letter, zeghidour2018fully}. %However, as accuracy is not our main objective, we kept our model architecture deliberately simple. % NEW
%Detailed description is beyond the scope of this paper. We give here the main principles. No pre-processing and no voice separation have been performed on the raw audio. %

\noindent\textbf{Model} It is an 8-layers 2D-CNN with 3x3 kernels. Max-pooling 2x2 is applied on the first 2 layers to decrease time and frequency dimensions, and max-pooling 1x2 is applied on the next 6 layers only for frequency dimension. The first layer has 64 filters, doubled at each layer up to 512. A dropout with rate 0.3 is applied.

\begin{figure}[!h]
%\vspace{-0.5em}
 \centerline{
 %\framebox{
    \includegraphics[width=1\columnwidth]{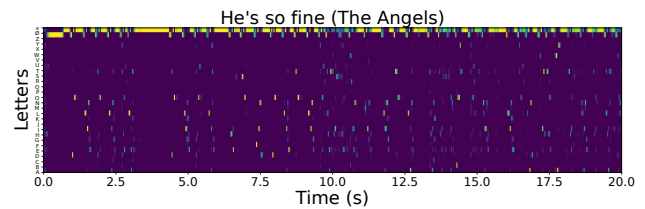}\\
    }
 %}
 %\vspace{-0.5em}
\caption{A posteriorgram obtained for 20 sec. of audio.}   
    \label{fig:ctc_outputs}
    %\vspace{-0.5 cm}
\end{figure}

\noindent\textbf{Inputs/outputs} We use no pre-processing on the audio (no data augmentation, no voice separation). As classically done in ASR, we use 40 band Mel-spectrogram with 10ms timeframes as input feature. The model outputs a posteriorgram corresponding to the log-probabilities of the 'a...z' letters, the space symbol and the CTC blank symbol, i.e. 28 bins in total. An output example is shown \figref{fig:ctc_outputs}. 

\noindent\textbf{Training and inference} The model is trained on the Dali dataset, which provide 12k+ polyphonic audio with lyrics annotations at the word level. We used audio chunks of 10 seconds, and used a CTC loss with Adam optimizer to train the model to align between audio and text. We evaluated its performances on a distinct Dali test subset using a greedy beam search decoding with no language model, and achieved a modest Character Error Rate (CER) of 0.496.

\subsection{Convolutional architecture}

The same simple architecture (with different configurations) has shown its ability to capture relevant melodic or harmonic similarity between versions \cite{doras2019cover, yesiler2019accurate}. It consists in a plain 5-layers Convolutional Neural Network (CNN), encoding each input feature using time and frequency max-pooling, while increasing the number of filters at each layer. The CNN output feeds a gated temporal attention mechanism \cite{serra2018towards}, which was found to help the model to focus on relevant portions of the input features. We refer the reader to our previous work \cite{doras2020combining} for implementation details.% NEW
%All the details of the architecture and the principles are provided in \cite{doras2020combining}, please refer to Figure 1 from \cite{doras2020combining} for an illustrated example of architecture. %

In this work, we keep the exact same generic architecture, and propose two new configurations to process also the rhythmic and lyrics features. We summarize the configuration used for each of the four features in \tabref{tab:cnn_model}. 

\noindent \textbf{Rhythmic features} Given the tempo invariance on the periodicity dimension explained in  \secref{subsec:rhythmic_features}, the first layer has a 20 bins kernel on this axis to capture all patterns within 2 octaves (for instance quarter, eight and sixteenth notes). All other layers have 3x3 kernel with max-pooling of size 2 on the periodicity axis.

\noindent \textbf{Lyrics features} We conducted various experiments to find the best kernel size to apply to the lyrics. It appeared that a rather short receptive field of 10 bins yield the best results. We kept it for all layers, with a mean-pooling of size 5.

Finally, a dense layer applied after the temporal attention block outputs a 512 bins embeddings, L2-normalized so that each track becomes a point on the surface of the unit hypersphere, bounding the distance between 2 points within the [0,2] interval.

\begin{table*}[t]
\centering
\resizebox{\textwidth}{!}{
\begin{tabular}{*{26}{c|}}
%& \multicolumn{3}{c |}{SHS$_4$}&\multicolumn{3}{c}{Da-TACOS}\\
Features    
&\multicolumn{6}{c |}{Melodic \cite{doras2019cover}}
&\multicolumn{6}{c |}{Harmonic \cite{yesiler2019accurate}}
&\multicolumn{6}{c |}{Rhythmic}
&\multicolumn{6}{c |}{Lyrics}\\
\hline
Layers
&n &k &ks &p &ps &d 
&n &k &ks &p &ps &d 
&n &k &ks &p &ps &d 
&n &k &ks &p &ps &d \\
\hline
1   
&64 &3x3 &1x1 &2x2 &2x2 &0.0    
&256 &180x12 &1x1 &1x12 &1x1 &0.0                                               
&64 &3x20 &1x1 &1x2 &1x2 &0.4 %r   
&64 &10 &1 &5 &2 &0.3  \\ %l   

2 
&128 &3x3 &1x1 &2x2 &2x2 &0.1     
&256 &5x1 &1x1 &1x1 &1x1 &0.0                                              
&128 &3x3 &1x1 &1x2 &1x2 &0.3    
&128 &10 &1 &5 &2 &0.2  \\ %l   

3
  
&256 &3x3 &1x1 &2x2 &2x2 &0.1                                               
&256 &5x1 &1x1 &1x1 &1x1* &0.0    
&256 &3x3 &1x1 &1x2 &1x2 &0.2                                                
&256 &10 &1 &5 &2 &0.1  \\ %l 

4
&512 &3x3 &1x1 &2x2 &2x2 &0.2      
&512 &5x1 &1x1 &1x1 &1x1 &0.0                                                  
&512 &3x3 &1x1 &1x2 &1x2 &0.1                                                 
&512 &10 &1 &5 &2 &0.1 \\  %l       \\

5
&1024 &3x3 &1x1 &2x2 &2x2 &0.3  
&512 &5x1 &1x1 &1x1 &1x1* &0.0       
&1024 &3x3 &1x1 &1x2 &1x2 &0.0                                                  
&1024 &10 &1 &5 &2 &0.0 \\ %l      \\
\end{tabular}
}
\caption{Configuration of the 5-layers CNN used for the features. n : number of filters, k : kernel size, ks : kernel stride, p : pool size, ps : pool stride, d : dropout. 1st dimension is time, 2d dimension is frequency. Convolutions are "same" for Me, Rh and Ly, and "valid" for Ha. *Ha uses dilation rate of 20 and 13 along time dimension on $3^{\text{rd}}$ and $5^{\text{th}}$ layers respectively.}
\label{tab:cnn_model}
%\vspace{-0.4 cm}
\end{table*}

\subsection{Embedding concatenation}
\label{sec:emb_concat}

In this work, we investigated only a late embedding fusion scheme, which simply consists in concatenating each feature embedding, and to L2-normalize the concatenated result. 
% NEW
%The main objective is to show the role of the lyrics, therefore we didn't investigate other fusion scheme. 
%
It is straightforward to show that the distance between a pair of normalized concatenation of $n$ feature embeddings is the quadratic mean of the $n$ distances between each feature embedding pair. All feature combination scores in \secref{sec:quantitative} have been obtained using this method.
 
 The practical advantage of this simple concatenated embedding is twofold: it remains easy to store and to query in an large index, and the lookup can be done for some specific feature combinations only (zero masking the unwanted feature embeddings).

\section{Experiment and results}
\label{sec:quantitative}

In this section, we present our experimental setup and results. For brevity, we will denote melodic, harmonic, rhythmic and lyrics features by their abbreviations Me, Ha, Rh, and Ly, respectively.
%\vspace{-0.3 cm}
\subsection{Experimental setup}
\label{subsec:exp_setup}

We use the exact same protocol as in our previous work \cite{doras2020combining}.

\noindent \textbf{Training} We trained our four models on the publicly available dataset \shsfive$^1$%\footnotemark[\ref{note1}]
, which contains \textasciitilde 62k covers of \textasciitilde 7.5k works. We used the provided features for Me and Ha, and extracted Rh and Ly from the audio, as described in sections \ref{subsec:rhythmic_features} and \ref{subsec:lyrics_features}. We used a semi-hard triplet loss, using an Adam optimizer, an initial learning rate of $1e^{-4}$ and a batch size of 64. All other details are replicated from \cite{doras2020combining}.

\noindent \textbf{Test} We tested our four models on \shsfour$^1$%\footnotemark[\ref{note1}]
, containing \textasciitilde 50k covers of \textasciitilde 20k works. We also retrained models on \shsfivefour and tested on Da-Tacos\footnote{https://github.com/MTG/da-tacos}, containing 13k covers of 1k works and 2k confusing tracks \cite{yesiler2019tacos}. 
As some samples overlap between Da-Tacos, \shsfour and Dali, we made sure that none of these samples were used for scoring the models, as done in \cite{du2022bytecover2}.

For each feature, we used the corresponding trained model to compute each track embedding, and computed their pairwise distance matrix. For feature combinations, distance matrix is computed using the quadratic mean of each feature distance matrix, as described in \secref{sec:emb_concat}.

\subsection{Results}
\label{subsec:results}

We summarize in \tabref{tab:averaging_scores_comparison} the performances obtained on \shsfour and Da-Tacos by our models, reporting the metrics classically used for VI: Mean Average Precison (MAP), mean number of correct answers in the first 10 (MT@10) and mean rank of first correct answer (MR1).

\begin{table}[!h]
%\vspace{-0.2 cm}
\centering
\resizebox{\columnwidth}{!}{
\begin{tabular}{l|c|c|c|c|c|c}
Train set  & \multicolumn{3}{c |}{\shsfive}&\multicolumn{3}{c}{Pruned \shsfivefour}\\
\hline
Test set  & \multicolumn{3}{c |}{Pruned \shsfour}&\multicolumn{3}{c}{Pruned Da-Tacos}\\
\hline
Input feature      &MAP    &MT@10   &MR1        &MAP        &MT@10       &MR1\\
\hline
Me       &0.427  &0.822   &1131        & 0.363      &4.064 &97     \\
Ha        &0.538  &1.003   &982        & \textbf{0.488}      &\textbf{5.256}      &\textbf{63}     \\
Rh        &0.099    &0.231     &2921        & 0.055  &0.689  &244    \\
Ly          & \textbf{0.672} & \textbf{1.190}  & \textbf{968}      & 0.393      & 4.596      & 199 \\
%Ly (de)      & \textbf{0.673}   &\textbf{1.273}   & \textbf{89}      &\textbf{0.808}     &\textbf{8.176}      &\textbf{8}     \\
\hline
Me+Ha       &0.693  &1.256   &453        & \textbf{0.626}      &\textbf{6.668}      &\textbf{32}     \\
%Ha+Ly       &0.778  &1.215   &369        &XXX      &XXX      &XXX     \\
%Ha+Ly (de)      &\textbf{0.825}  &\textbf{1.317}   &\textbf{19}        &XXX      &XXX      &XXX     \\
%Me+Ly       &0.728  &1.141   &512        &XXX      &XXX      &XXX     \\
%Me+Ly (de)      &0.761  &1.255   &41        &XXX      &XXX      &XXX     \\
%Me+Rh      &0.456  &0.790   &1264        &XXX      &XXX      &XXX     \\
%Rh+Ly      &0.585  &0.955   &693        &XXX      &XXX      &XXX     \\
%Rh+Ly (de)     &0.511  &0.970   &57        &XXX      &XXX      &XXX     \\
%Ha+Rh      &0.564  &0.951   &1070        &XXX      &XXX      &XXX     \\
\hline
Me+Ha+Ly  & \textbf{0.800}  &\textbf{1.396}   &291        &0.602      &6.480      &33     \\
Me+Ha+Rh  &0.688  &1.250 &413        &0.557 &5.994      &33     \\
%Me+Ha+Ly (de)  &\textbf{0.888}  &\textbf{1.383}   &\textbf{15}        &\textbf{0.874}      &\textbf{8.870}      &\textbf{2}     \\
%Me+Rh+Ly  &0.699  &1.110   &451        &XXX      &XXX      &XXX     \\
%Ha+Rh+Ly  &0.757  &1.194   &356        &XXX      &XXX      &XXX     \\
\hline

%Me+Ha+Rh+Ly      &0.800  &1.250   &306        &XXX      &XXX      &XXX     \\
Me+Ha+Rh+Ly     &0.785  &1.378   &\textbf{286}       &0.560 &6.054      &33     \\
\hline
\hline
%\hline
Me+Ha (O)   &0.879  &1.521   &97         &0.837 &8.709     &4    \\
%Ha+Ly (O)    &0.929  &1.409   &104     &XXX &XXX      &XXX \\
%Ha+Ly (de) (O)    &\textbf{0.998}  &\textbf{1.491}   &\textbf{1.11}     &XXX &XXX      &XXX \\
%Me+Ly (O)    &0.883  &1.349   &101     &XXX &XXX      &XXX     \\
%Me+Ly (de) (O)    &\textbf{0.998}  &1.490   &2     &XXX &XXX      &XXX     \\
%Me+Rh (O)    &0.750  &1.195   &173     &XXX &XXX      &XXX \\
%(f)Rh+Ly (O)    &0.853  &1.316   &118     &XXX &XXX      &XXX \\
%Rh+Ly (de) (O)    &0.995  &1.488   &2     &XXX &XXX      &XXX \\
%Ha+Rh (O)    &0.824  &1.285   &128     &XXX &XXX      &XXX \\
Me+Ha+Rh (O)    &0.939  &1.607   &21     &0.905 &9.303     &\textbf{1}  \\
Me+Ha+Ly (O)    &\textbf{0.963}  &\textbf{1.637}   &\textbf{14}     &\textbf{0.918} &\textbf{9.398}      &\textbf{1} \\
%Me+Ha+Ly (de) (O)    &\textbf{0.999}  &\textbf{1.493}   &\textbf{1}     &\textbf{0.999} &\textbf{9.984}      &\textbf{1} \\
%Me+Rh+Ly (O)    &0.940  &1.424   &23     &XXX &XXX      &XXX \\
%Me+Rh+Ly (de) (O)    &\textbf{0.999}  &\textbf{1.493}   &1.27     &XXX &XXX      &XXX \\
%Ha+Rh+Ly (O)    &0.962  &1.451   &16    &XXX &XXX      &XXX \\
%Ha+Rh+Ly (de) (O)    &\textbf{0.999}  &\textbf{1.493}   &\textbf{1.00...}    &XXX &XXX      &XXX \\
\hline
%Me+Ha+Rh+Ly (O) &0.980  &1.470   &5     &XXX &XXX      &XXX \\
Me+Ha+Rh+Ly (O) &\textbf{0.978} &\textbf{1.658}   &\textbf{4}     &\textbf{0.951}      &\textbf{9.657}      &\textbf{1}
\end{tabular}
}
\caption{Performance metrics obtained on \shsfour and Da-Tacos for input features and their combinations. O=oracle}
%\vspace{-0.8em}
\label{tab:averaging_scores_comparison}
\end{table}

\noindent We first examine the performances of each single feature. Me and Ha results are in line with our previous work \cite{doras2020combining}. %(Me results are slightly better here because we found out that using a temporal resolution of 46ms for melody extraction improved the results over the original 93ms resolution)

\noindent \textbf{Rhythmic features} The performances obtained using Rh are clearly lower, which suggests that our rhythm feature is not providing relevant VI information, or that the rhythm patterns themselves are not specific enough to discriminate between versions and non-versions. As a consequence, adding Rh degrades our Me+Ha baseline. This is not entirely surprising, as many non-versions might exhibit similar rhythmic patterns (we will however see in \secref{subsec:oracle} and \secref{sec:qualitative} that Rh can be relevant in some cases).  

\noindent \textbf{Lyrics features} In contrast, the use of lyrics clearly outperforms the other features on \shsfour. This confirms that versions often share the same lyrics, and that even a very inaccurate ALR system can be beneficial to VI. The combination Me+Ha+Ly improves by more than 10\% the Me+Ha performances, which will be explained in \secref{sec:qualitativeMeHaLy}. 

On Da-Tacos, we observe much less clear-cut results. As already noticed by Vaglio et al.\cite{vaglio2020audio}, Da-Tacos has about 20\% of instrumental songs. A closer look to our results shows that these songs typically produce false positives. Following Vaglio et al., we pruned the instrumental songs from Da-Tacos, and recomputed the results for Ly, obtaining MAP=0.674, MT@10=5.931 and MR1=59, which is consistent with the values obtained for \shsfour.

\subsection{Oracle results}
\label{subsec:oracle}

We also present in \tabref{tab:averaging_scores_comparison} the results obtained by an oracle. This oracle only considers the best performing feature to compute each pairwise distance: for each pair of tracks, it only uses the feature embeddings yielding the lowest (resp. highest) distance for versions (resp. non-versions).

Interestingly, it appears that the performances of an oracle using three or four features approaches the theoretical optimal values on our two datasets. This is particularly clear with the Me+Ha+Ly and Me+Ha+Rh+Ly combinations: the   MAP tends to $1.0$, i.e. most versions have been ranked in the first answers for each query. The MR1 also tends to $1$, i.e. first answer is correct for most queries. The MT@10 also tends to the theoretical optimal values (in \shsfour, each track has 2, 3 or 4 versions, and its optimal MT@10=1.695, while in Da-Tacos, each track has 0 or 12 versions, and its optimal MT@10=10).

\begin{figure}[h!]
\vspace{-0.5 em}
    \centering{
        \includegraphics[width=.95\columnwidth]{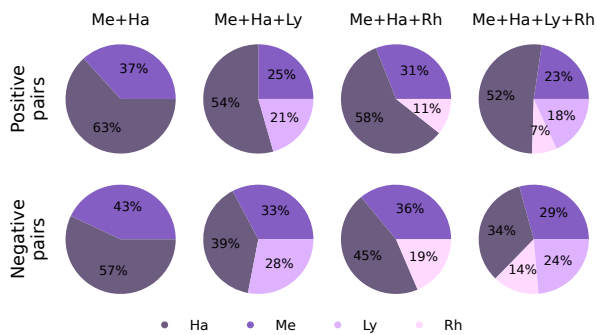}
        }
    \vspace{-0.3 cm}
    \caption{Most relevant feature proportions to identify positive and negative pairs on \shsfour.}   
    \label{fig:oracle_stats}
    \vspace{-0.5 em}
\end{figure}

The \figref{fig:oracle_stats} displays the contribution of each feature to the oracle score, i.e. which feature is the most relevant to determine if two tracks are versions ("Positive pairs") or non-versions ("Negative pairs"). 

It appears that Ha is usually the most efficient feature to discriminate between versions or non-versions. However, both Me and Ly contribute importantly to identification (e.g. resp. 25\% and 21\% of the positive pairs, 33\% and 28\% of the negative pairs in the Me+Ha+Ly combination). Finally, and even though Rh alone yields poor results, its contribution is not negligible when combined with other features (e.g. 14\% of the negative pairs in the Me+Ha+Rh+Ly combination).

\subsection{Comparison with state-of-the-art}

Performances obtained by recent VI systems are summarized in \tabref{tab:soa_comparison} (second column indicates the embedding size used by each system).

Our system using Me+Ha+Ly improves the state-of-the-art on \shsfour and on Da-Tacos (without instrumentals).

\begin{table}[!h]
\centering
\resizebox{\columnwidth}{!}{
\begin{tabular}{l|c|c|c|c|c|c|c}
Test set &  & \multicolumn{3}{c |}{\shsfour}& \multicolumn{3}{c}{Da-Tacos}\\
\hline
Model  &Emb.   &MAP    &MT@10   &MR1        &MAP        &MT@10       &MR1\\
\hline

Doras et al. \cite{doras2020combining} &512  &0.660 &1.080  &657  &0.635 &6.744  &30  \\
Vaglio et al. \cite{vaglio2020audio} &n/a    &n/a    &n/a    & n/a    &0.804* &n/a  &n/a\\
Du et al. \cite{du2022bytecover2} &  1536          &n/a    &n/a    &n/a   &\textbf{0.791} &n/a     &\textbf{19.2}\\
\hline
\multirow{2}{*}{Me+Ha+Ly (ours)} &\multirow{2}{*}{1536}  &\multirow{2}{*}{\textbf{0.800}}  &\multirow{2}{*}{\textbf{1.396}}   &\multirow{2}{*}{\textbf{291}}  &\textbf{0.818*} & \textbf{7.205*} & \textbf{16*} \\ &  & &  & &  0.602     &6.480      &33 

% Me+Ha+Ly (ours)  &1536 &\textbf{0.888}  &\textbf{1.383}   &\textbf{15}        &  0.606     &6.512      &33  
% \cline{6-8}
% \textbf{0.821} & \textbf{7.234} & \textbf{16}
\end{tabular}
}
\caption{Sota comparison on \shsfour and Da-Tacos. $^*$results obtained on Da-Tacos-Vocals (w/o instrumental tracks).}
%\vspace{-0.8em}
\label{tab:soa_comparison}
\vspace{-0.5em}
\end{table}

%\FloatBarrier
\section{Qualitative analysis}
\label{sec:qualitative}

In this section, we illustrate qualitatively the previous quantitative results. We encourage readers to listen to the audio samples available on the paper companion website$^1$%\footnotemark[\ref{note1}]  
\hspace{0.001 cm} as part of reading the paper.

\subsection{Ly vs. Me+Ha examples}
\label{sec:qualitativeMeHaLy}
\begin{figure}[h!]
%\vspace{-0.3 cm}
    \centering{
        \includegraphics[width=0.7\columnwidth]{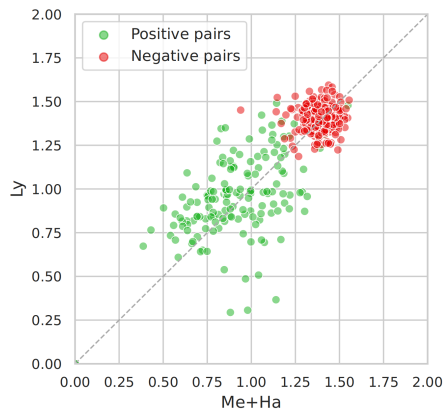}
        }
        \vspace{-0.2 cm}
    \caption{Pairwise distances for Me+Ha vs. Ly (500 pairs from \shsfour).}   
    \label{fig:cluster}
    \vspace{-0.3 cm}
\end{figure}

We intend here to illustrate how Ly improves the Me+Ha system. \figref{fig:cluster} plots the distance obtained for Me+Ha and Ly between randomly sampled positive (green) and negative (red) pairs.

\noindent This plot confirms that Me+Ha and Ly are complementary, as already seen in \secref{subsec:oracle}. The dots situated away from the diagonal correspond to track pairs scored correctly by Me+Ha and incorrectly by Ly (or vice-versa). As certain features yield better results for certain songs, their combination will statistically improve the results, as we will now illustrate with some contrasted examples.

\begin{figure*}[t]
         \begin{subfigure}[b]{0.5\textwidth}
        \includegraphics[width=\textwidth]{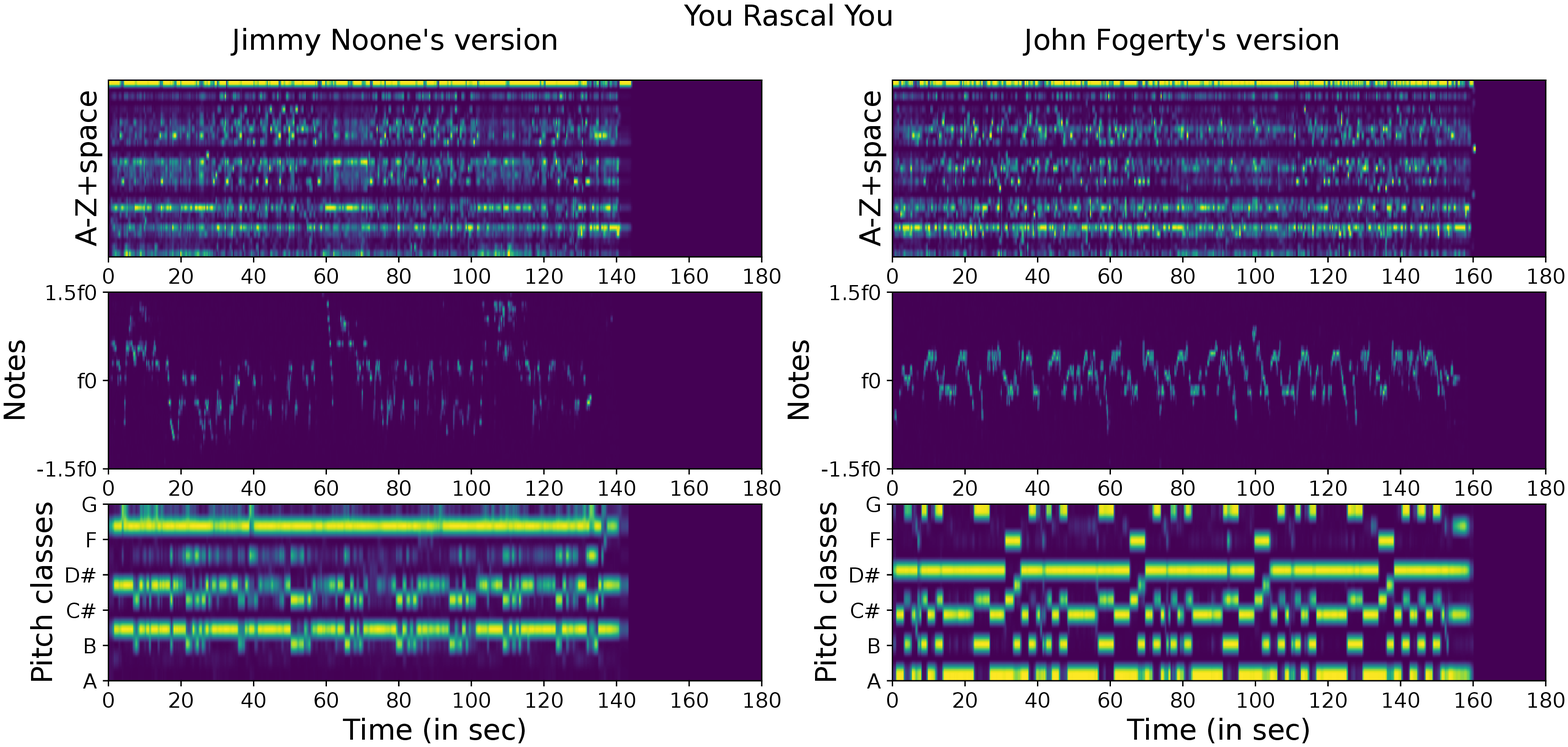}
        \caption{Ly better than Me+Ha-  d$_{\text{Me+Ha}}$=1.470, d$_{\text{Ly}}$=0.238}
        \label{fig:ly_illus}
    \end{subfigure}
    \hfill
    \begin{subfigure}[b]{0.5\textwidth}
        \includegraphics[width=\textwidth]{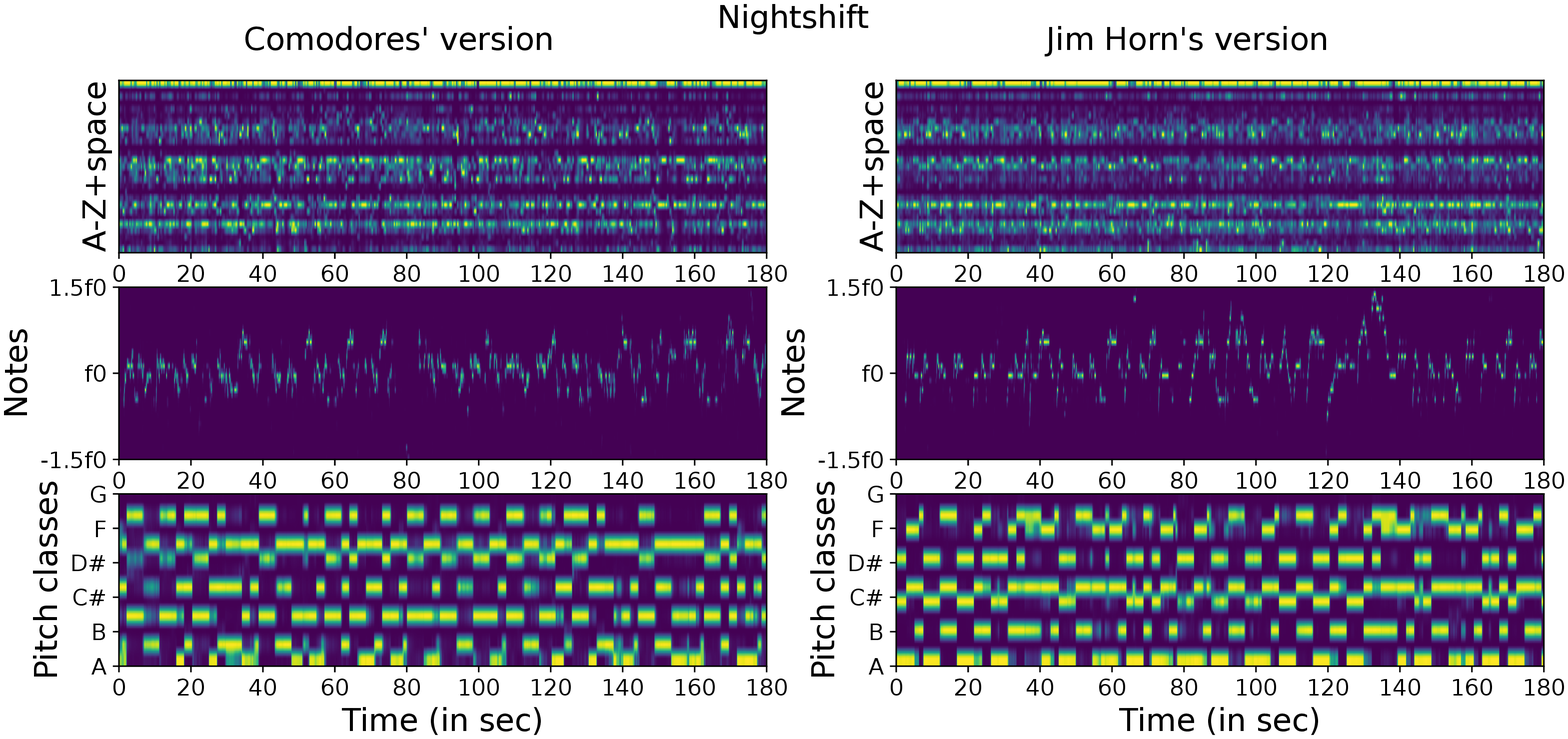}
        \caption{Ly worse than Me+Ha-  d$_{\text{Me+Ha}}$=0.597, d$_{\text{Ly}}$=1.282}
        \label{fig:ly_contras}
    \end{subfigure}
    \begin{subfigure}[b]{0.5\textwidth}
        \includegraphics[width=\textwidth]{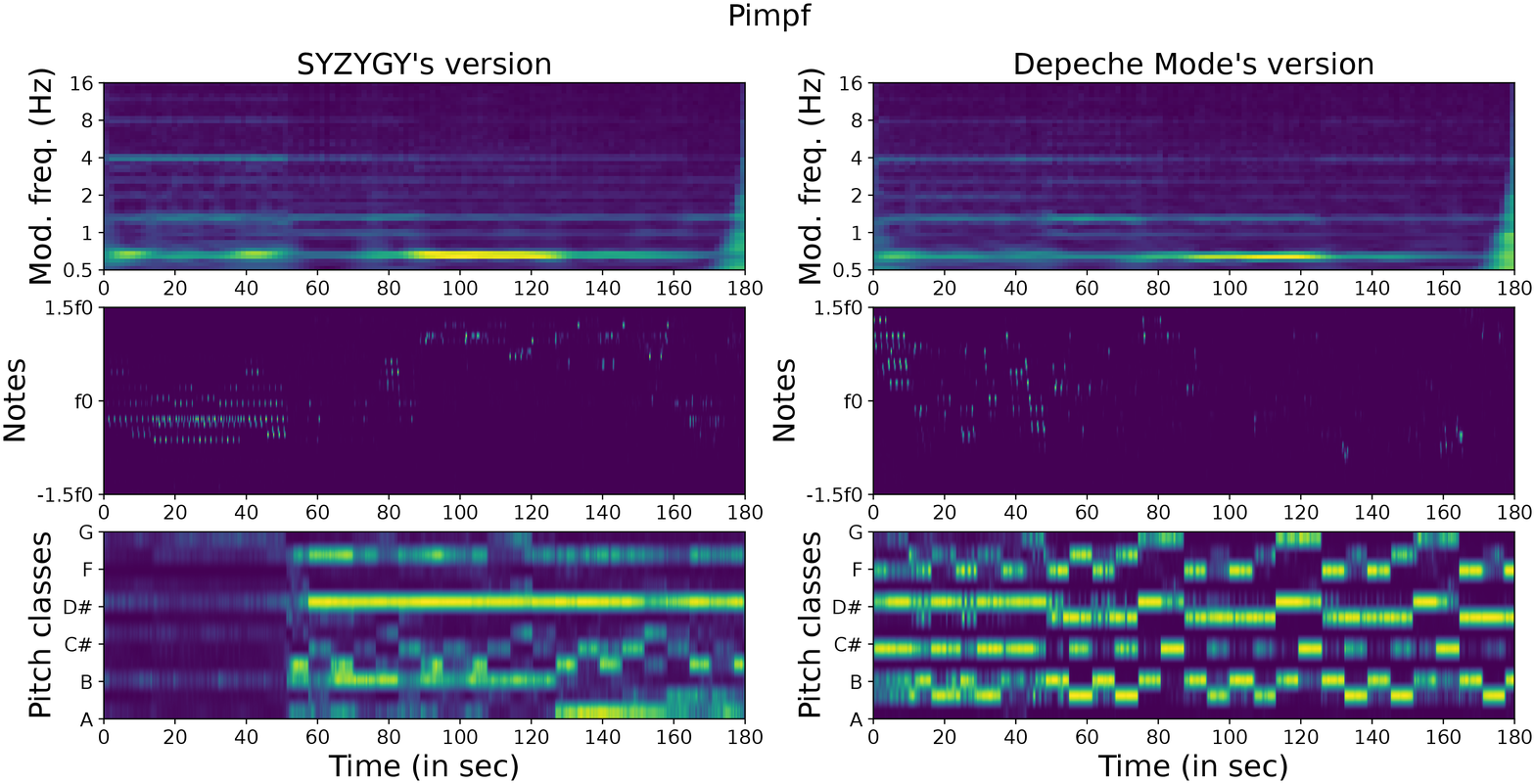}
        \caption{Rh better than Me+Ha -  d$_{\text{Me+Ha}}$=1.084, d$_{\text{Rh}}$=0.459}
        \label{fig:fp_illus}
    \end{subfigure}
    \hfill
    \begin{subfigure}[b]{0.5\textwidth}
        \includegraphics[width=\textwidth]{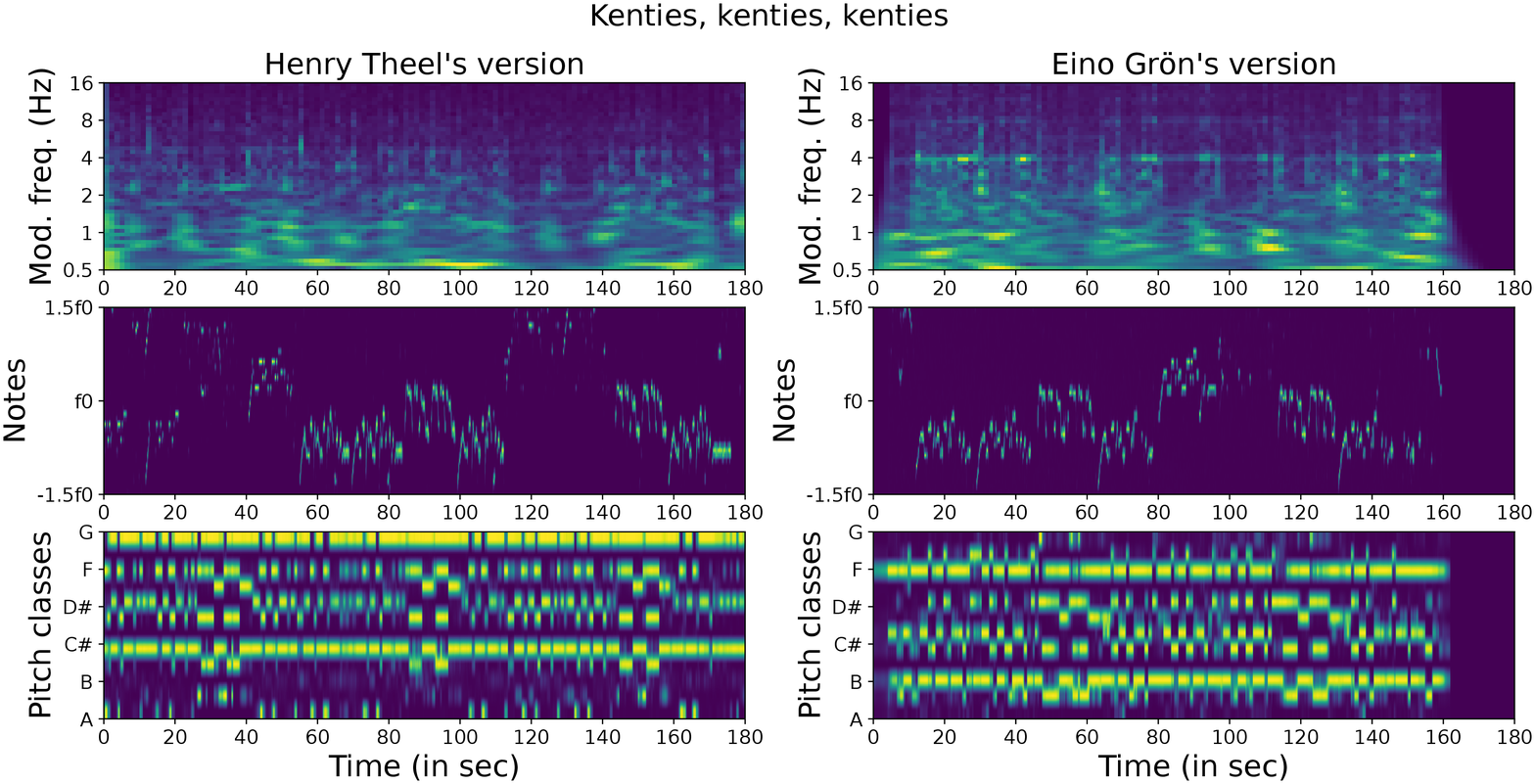}
        \caption{Rh worse than Me+Ha -  d$_{\text{Me+Ha}}$=0.498, d$_{\text{Rh}}$=1.380}
        \label{fig:fp_contras}
    \end{subfigure}
    \label{fig:illus}
    \vspace{-0.5 cm}
    \caption{Examples for rhythm and lyrics : Rh > Me+Ha \ref{fig:fp_illus}, Rh < Me+Ha \ref{fig:fp_contras}, Ly > Me+Ha \ref{fig:ly_illus} and Ly < Me+Ha \ref{fig:ly_contras}} 
    \vspace{-0.3 cm}
\end{figure*}

\noindent \textbf{Ly > Me+Ha} There are many versions whose musical style, melody and harmony differ greatly from the original, and where only the lyrics can help to identify them. This is illustrated on \figref{fig:ly_illus}, which shows that the Jimmy Noone's and John Fogerty's versions of "You Rascal You" are very different musically while the lyrics exhibit enough similarity to be correctly identified. 

% \begin{figure}[h!]
%     \includegraphics[width=0.45\textwidth]{figs/you_rascal_you.pdf}
%     \caption{Illustrative example for lyrics -  $d_{Me+Ha}=1.470$, $d_{Ly}=0.037$}   
%     \label{fig:ly_illus}
% \end{figure}

It also appears that our approximated ALR system is efficient for different languages. For instance, the versions of Asta Kask and of The Hep Stars of the song "I natt jag drömde", are very different in melody and harmony (d$_{\text{Me+Ha}}$=1.448) while the lyrics remain similar (d$_{\text{Ly}}$=0.342), despite the fact that the lyrics are in Swedish.

\noindent \textbf{Ly < Me+Ha} As already mentioned, using Ly will yield wrong results in presence of instrumental version (i.e. no lyrics). In the example shown in \figref{fig:ly_contras}, the version of "Nightshift" by the Commodores has lyrics while the one of Jim Horn's does not. We noticed that the system sometimes considers lead instruments as voices. However, this Ly false negative is correctly caught by the Me+Ha. 

\subsection{Rh vs. Me+Ha examples}

Although less obvious than for Ly, combining Rh and Me+Ha also appears to be complementary in some cases.

\noindent \textbf{Rh > Me+Ha} Even though Rh yields poor performances in general, there are cases where it is the only feature available to identify versions. This is illustrated on \figref{fig:fp_illus}, which shows the Rh, Me and Ha features for two versions of "Pimpf". In this song, the melody is almost non-existent, and the harmony is very different between both versions. Only a few bass notes in the middle are salient enough to identify the song, and this short bassline appears similarly on the two FP features.

Rh might also be a good discriminating feature in other cases, e.g. for live concert versions. One version of "Mama's Little Baby" is recorded in studio while the other is a concert filmed from the audience. The Me+Ha distance between these versions is high (d$_{\text{Me+Ha}}$=1.329) because of the bad live recording quality. But, the drums are distinguishable enough to find similarity (d$_{\text{Rh}}$=0.714).

\noindent \textbf{Rh < Me+Ha} But Rh often yields wrong results. This is illustrated on \figref{fig:fp_contras}, which shows the features of two versions of "Kenties kenties kenties". Although melody and harmony are similar, the rhythm is very different, and the use of Rh produces a false negative.
%Another example is "Doesn't Anybody Know My Name", which does not exhibit the same rhythmic pattern in Hank Williams' version as in Tommy Roe's (d$_{\text{Me+Ha}}$=0.526, d$_{\text{Rh}}$=1.476).

%\subsection{Ly vs. Me+Ha examples}

% \begin{figure}[h!]
%     \includegraphics[width=0.47\textwidth]{figs/nightshift.pdf}
%     \caption{Constrative example for lyrics -  $d_{Me+Ha}=0.597$, $d_{Ly}=1.721$}   
%     \label{fig:ly_contras}
% \end{figure}

\vspace{-0.5em}
\section{Conclusion}

It was shown previously that VI systems combining melody and harmony yields promising performances. In this paper, we proposed to consider also rhythmic and lyrics features to improve these results further. We showed that an existing rhythmic feature commonly used for genre classification is only helpful in a few cases, such as live version identification. But we also showed that an approximate lyrics representation can improve the performances of existing melody and harmony-based systems. We explained these results by the fact that detecting correctly only a few character sequences appears to be enough to distinguish versions and non-versions. We showed that our system combining these features establishes new state-of-the-art on two public datasets. More importantly, we indicated that these feature combinations provide enough information to approach the theoretical optimal performances obtainable on these datasets.

In our future work, we will investigate a more elaborated fusion scheme in order to train our model to behave as an oracle: our objective is to teach the system how to choose between available features to pick only the most relevant one for each pair of tracks. This might answer the question of whether the concept of musical version can be reduced to its melodic, harmonic, rhythmic, and lyrics dimensions.

\section{Acknowlegments}

Authors would like to thank the anonymous reviewers for their helpful comments, and for suggesting the \figref{fig:oracle_stats}.

% For bibtex users:
\bibliography{thesis}

\begin{thebibliography}{10}
\providecommand{\url}[1]{#1}
\csname url@samestyle\endcsname
\providecommand{\newblock}{\relax}
\providecommand{\bibinfo}[2]{#2}
\providecommand{\BIBentrySTDinterwordspacing}{\spaceskip=0pt\relax}
\providecommand{\BIBentryALTinterwordstretchfactor}{4}
\providecommand{\BIBentryALTinterwordspacing}{\spaceskip=\fontdimen2\font plus
\BIBentryALTinterwordstretchfactor\fontdimen3\font minus
  \fontdimen4\font\relax}
\providecommand{\BIBforeignlanguage}[2]{{%
\expandafter\ifx\csname l@#1\endcsname\relax
\typeout{** WARNING: IEEEtran.bst: No hyphenation pattern has been}%
\typeout{** loaded for the language `#1'. Using the pattern for}%
\typeout{** the default language instead.}%
\else
\language=\csname l@#1\endcsname
\fi
#2}}
\providecommand{\BIBdecl}{\relax}
\BIBdecl

\bibitem{yesiler2021audio}
F.~Yesiler, G.~Doras, R.~M. Bittner, C.~J. Tralie, and J.~Serr{\`a},
  ``Audio-based musical version identification: Elements and challenges,''
  \emph{IEEE Signal Processing Magazine}, vol.~38, no.~6, 2021.

\bibitem{yu2019learning}
Z.~Yu, X.~Xu, X.~Chen, and D.~Yang, ``Learning a representation for cover song
  identification using convolutional neural network,'' \emph{Proceedings of
  ICASSP (International Conference on Acoustics, Speech and Signal
  Processing)}, 2020.

\bibitem{doras2019cover}
G.~Doras and G.~Peeters, ``Cover detection using dominant melody embeddings,''
  in \emph{Proceedings of ISMIR (International Society for Music Information
  Retrieval)}, 2019.

\bibitem{yesiler2019accurate}
F.~Yesiler, J.~Serr{\`a}, and E.~G{\'o}mez, ``Accurate and scalable version
  identification using musically-motivated embeddings,'' \emph{Proceedings of
  ICASSP (International Conference on Acoustics, Speech and Signal
  Processing)}, 2020.

\bibitem{du2022bytecover2}
X.~Du, K.~Chen, Z.~Wang, B.~Zhu, and Z.~Ma, ``Bytecover2: Towards
  dimensionality reduction of latent embedding for efficient cover song
  identification,'' in \emph{ICASSP 2022-2022 IEEE International Conference on
  Acoustics, Speech and Signal Processing (ICASSP)}.\hskip 1em plus 0.5em minus
  0.4em\relax IEEE, 2022, pp. 616--620.

\bibitem{lee2020disentangled}
J.~Lee, N.~J. Bryan, J.~Salamon, Z.~Jin, and J.~Nam, ``Disentangled
  multidimensional metric learning for music similarity,'' in \emph{ICASSP
  2020-2020 IEEE International Conference on Acoustics, Speech and Signal
  Processing (ICASSP)}.\hskip 1em plus 0.5em minus 0.4em\relax IEEE, 2020.

\bibitem{yu2019temporal}
Z.~Yu, X.~Xu, X.~Chen, and D.~Yang, ``Temporal pyramid pooling convolutional
  neural network for cover song identification,'' \emph{Proceedings of the
  International Joint Conference on Artificial Intelligence}, 2019.

\bibitem{doras2020combining}
G.~Doras, F.~Yesiler, J.~Serr{\`a}, E.~Gomez, and G.~Peeters, ``Combining
  musical features for cover detection,'' in \emph{Proceedings of ISMIR
  (International Society for Music Information Retrieval)}, 2020.

\bibitem{vaglio2021words}
A.~Vaglio, R.~Hennequin, M.~Moussallam, and G.~Richard, ``The words remain the
  same - cover detection with lyrics transcription,'' in \emph{Proceedings of
  ISMIR (International Society for Music Information Retrieval)}, 2021.

\bibitem{douwes2021energy}
C.~Douwes, P.~Esling, and J.-P. Briot, ``Energy consumption of deep generative
  audio models,'' \emph{arXiv preprint arXiv:2107.02621}, 2021.

\bibitem{slaney2008learning}
M.~Slaney, K.~Weinberger, and W.~White, ``Learning a metric for music
  similarity,'' in \emph{International Symposium on Music Information Retrieval
  (ISMIR)}, vol. 148, 2008.

\bibitem{mcfee2012learning}
B.~McFee, L.~Barrington, and G.~Lanckriet, ``Learning content similarity for
  music recommendation,'' \emph{IEEE Transactions on Audio, Speech, and
  Language Processing}, vol.~20, no.~8, pp. 2207--2218, 2012.

\bibitem{schroff2015facenet}
F.~Schroff, D.~Kalenichenko, and J.~Philbin, ``Facenet: A unified embedding for
  face recognition and clustering,'' in \emph{Proceedings of IEEE CVPR
  (Conference on Computer Vision and Pattern Recognition)}, 2015.

\bibitem{foote2002audio}
J.~Foote, M.~Cooper, and U.~Nam, ``Audio retrieval by rhythmic similarity.'' in
  \emph{ISMIR}.\hskip 1em plus 0.5em minus 0.4em\relax Citeseer, 2002.

\bibitem{dixon2004towards}
S.~Dixon, F.~Gouyon, G.~Widmer \emph{et~al.}, ``Towards characterisation of
  music via rhythmic patterns.'' in \emph{ISMIR}, 2004.

\bibitem{pampalk2006computational}
E.~Pampalk, ``Computational models of music similarity and their application in
  music information retrieval,'' Ph.D. dissertation, 2006.

\bibitem{pohle2009rhythm}
T.~Pohle, D.~Schnitzer, M.~Schedl, P.~Knees, and G.~Widmer, ``On rhythm and
  general music similarity.'' in \emph{ISMIR}, 2009, pp. 525--530.

\bibitem{foroughmand2019deep}
H.~Foroughmand and G.~Peeters, ``Deep-rhythm for tempo estimation and rhythm
  pattern recognition,'' in \emph{International Society for Music Information
  Retrieval (ISMIR)}, 2019.

\bibitem{trentin2001survey}
E.~Trentin and M.~Gori, ``A survey of hybrid ann/hmm models for automatic
  speech recognition,'' \emph{Neurocomputing}, vol.~37, no. 1-4, pp. 91--126,
  2001.

\bibitem{graves2006connectionist}
A.~Graves, S.~Fern{\'a}ndez, F.~Gomez, and J.~Schmidhuber, ``Connectionist
  temporal classification: labelling unsegmented sequence data with recurrent
  neural networks,'' in \emph{Proceedings of the 23rd international conference
  on Machine learning}.\hskip 1em plus 0.5em minus 0.4em\relax ACM, 2006, pp.
  369--376.

\bibitem{collobert2016wav2letter}
R.~Collobert, C.~Puhrsch, and G.~Synnaeve, ``Wav2letter: an end-to-end
  convnet-based speech recognition system,'' \emph{arXiv preprint
  arXiv:1609.03193}, 2016.

\bibitem{meseguer2018dali}
G.~Meseguer-Brocal, A.~Cohen-Hadria, and G.~Peeters, ``Dali: a large dataset of
  synchronized audio, lyrics and notes, automatically created using
  teacher-student machine learning paradigm.'' in \emph{19th International
  Society for Music Information Retrieval Conference}, ISMIR, Ed., 2018.

\bibitem{gupta2019acoustic}
C.~Gupta, E.~Y{\i}lmaz, and H.~Li, ``Acoustic modeling for automatic
  lyrics-to-audio alignment,'' \emph{arXiv preprint arXiv:1906.10369}, 2019.

\bibitem{stoller2019end}
D.~Stoller, S.~Durand, and S.~Ewert, ``End-to-end lyrics alignment for
  polyphonic music using an audio-to-character recognition model,'' \emph{arXiv
  preprint arXiv:1902.06797}, 2019.

\bibitem{gupta2020automatic}
C.~Gupta, E.~Y{\i}lmaz, and H.~Li, ``Automatic lyrics alignment and
  transcription in polyphonic music: Does background music help?'' in
  \emph{ICASSP 2020-2020 IEEE International Conference on Acoustics, Speech and
  Signal Processing (ICASSP)}.\hskip 1em plus 0.5em minus 0.4em\relax IEEE,
  2020, pp. 496--500.

\bibitem{wang2014improving}
X.~Wang and Y.~Wang, ``Improving content-based and hybrid music recommendation
  using deep learning,'' in \emph{Proceedings of the 22nd ACM international
  conference on Multimedia}, 2014, pp. 627--636.

\bibitem{zeghidour2018fully}
N.~Zeghidour, Q.~Xu, V.~Liptchinsky, N.~Usunier, G.~Synnaeve, and R.~Collobert,
  ``Fully convolutional speech recognition,'' \emph{arXiv preprint
  arXiv:1812.06864}, 2018.

\bibitem{serra2018towards}
J.~Serr{\`a}, S.~Pascual, and A.~Karatzoglou, ``Towards a universal neural
  network encoder for time series.'' in \emph{CCIA}, 2018.

\bibitem{yesiler2019tacos}
F.~Yesiler, C.~Tralie, A.~A. Correya, D.~F. Silva, P.~Tovstogan, E.~G{\'o}mez,
  and X.~Serra, ``Da-tacos: A dataset for cover song identification and
  understanding,'' in \emph{Proceedings of ISMIR (International Society for
  Music Information Retrieval)}, 2019.

\bibitem{vaglio2020audio}
A.~Vaglio, R.~Hennequin, M.~Moussallam, G.~Richard, and F.~d'Alch{\'e} Buc,
  ``Audio-based detection of explicit content in music,'' in \emph{ICASSP
  2020-2020 IEEE International Conference on Acoustics, Speech and Signal
  Processing (ICASSP)}.\hskip 1em plus 0.5em minus 0.4em\relax IEEE, 2020, pp.
  526--530.

\end{thebibliography}
% For non bibtex users:
%\begin{thebibliography}{citations}
% \bibitem{Author:17}
% E.~Author and B.~Authour, ``The title of the conference paper,'' in {\em Proc.
% of the Int. Society for Music Information Retrieval Conf.}, (Suzhou, China),
% pp.~111--117, 2017.
%
% \bibitem{Someone:10}
% A.~Someone, B.~Someone, and C.~Someone, ``The title of the journal paper,''
%  {\em Journal of New Music Research}, vol.~A, pp.~111--222, September 2010.
%
% \bibitem{Person:20}
% O.~Person, {\em Title of the Book}.
% \newblock Montr\'{e}al, Canada: McGill-Queen's University Press, 2021.
%
% \bibitem{Person:09}
% F.~Person and S.~Person, ``Title of a chapter this book,'' in {\em A Book
% Containing Delightful Chapters} (A.~G. Editor, ed.), pp.~58--102, Tokyo,
% Japan: The Publisher, 2009.
%
%
%\end{thebibliography}

\end{document}